\documentclass[twocolumn,floatfix,showpacs,prl,aps,tightenlines,superscriptaddress]{revtex4}
\usepackage{graphicx}
\usepackage{psfrag}
\usepackage{dcolumn}
\usepackage{bm}

\newcommand{\bea}{\begin{eqnarray}}
\newcommand{\eea}{\end{eqnarray}}
\newcommand{\beq}{\begin{equation}}
\newcommand{\eeq}{\end{equation}}
\newcommand{\KMS}{\rm km\,s^{-1}}
\newcommand{\vz}{v_\|}

\begin{document}

\title{Maximum gravitational recoil}
\author{Manuela Campanelli}
\affiliation{Center for Computational Relativity and Gravitation,
School of Mathematical Sciences,
Rochester Institute of Technology, 78 Lomb Memorial Drive, Rochester,
 New York 14623}
\affiliation{Center for Gravitational Wave Astronomy, Department of Physics and Astronomy,
The University of Texas at Brownsville, Brownsville, Texas 78520}

\author{Carlos O. Lousto} \affiliation{Center for Gravitational Wave Astronomy, Department of Physics and Astronomy,
The University of Texas at Brownsville, Brownsville, Texas 78520}
\affiliation{Center for Computational Relativity and Gravitation,
School of Mathematical Sciences,
Rochester Institute of Technology, 78 Lomb Memorial Drive, Rochester,
 New York 14623}

\author{Yosef Zlochower} \affiliation{Center for Gravitational Wave Astronomy,
Department of Physics and Astronomy, The University of Texas at Brownsville, Brownsville, Texas 78520}
\affiliation{Center for Computational Relativity and Gravitation,
School of Mathematical Sciences,
Rochester Institute of Technology, 78 Lomb Memorial Drive, Rochester,
 New York 14623}

\author {David Merritt} 
\affiliation{Department of Physics, 85 Lomb Memorial Drive, 
Rochester Institute of Technology, Rochester, NY 14623}
\affiliation{Center for Computational Relativity and Gravitation,
School of Mathematical Sciences,
Rochester Institute of Technology, 78 Lomb Memorial Drive, Rochester,
 New York 14623}

\date{\today}

\begin{abstract}
Recent calculations of gravitational radiation recoil generated during
black-hole binary mergers have reopened the possibility that a merged
binary can be ejected even from the nucleus of a massive host galaxy.
Here we report the first systematic study of gravitational recoil 
of equal-mass binaries with equal, but counter-aligned, spins 
parallel to the orbital plane.
Such an orientation of the spins is expected to maximize the recoil.
We find that recoil velocity (which is perpendicular to the 
orbital plane) varies sinusoidally with the angle
that the initial spin directions make with the initial linear momenta
of each hole and scales up to a maximum of $\sim 4000\ \KMS$ for
maximally-rotating holes. 
Our results show that the amplitude of the recoil velocity can depend 
sensitively on spin orientations of the black holes prior to merger.
\end{abstract}

\pacs{04.25.Dm, 04.25.Nx, 04.30.Db, 04.70.Bw} \maketitle

{\it Introduction:}
Generic black-hole-binary mergers will display a rich spectrum of
gravitational effects in the last few orbits prior to the formation of
the single rotating remnant hole.
These effects include spin and orbital plane precession, radiation of mass,
linear and angular momentum, as well as spin-flips of the remnant horizon.
Thanks to recent breakthroughs
in the full non-linear numerical evolution of black-hole-binary
spacetimes~\cite{Pretorius:2005gq,Campanelli:2005dd,Baker:2005vv}, it is
now possible to accurately simulate the merger process and examine these
effects in this highly non-linear regime~\cite{Campanelli:2006gf,
Baker:2006yw,Campanelli:2006uy,Campanelli:2006fg,Campanelli:2006fy,
Pretorius:2006tp,Pretorius:2007jn,Baker:2006ha,Bruegmann:2006at,
Buonanno:2006ui,Baker:2006kr,Scheel:2006gg,Baker:2007fb,Marronetti:2007ya,
Pfeiffer:2007yz}. 
Black-hole
binaries will radiate between $2\%$ and $8\%$ of their total mass and up to
$40\%$ of their angular momenta, depending on the magnitude and direction of
the  spin components, during the
merger~\cite{Campanelli:2006uy,Campanelli:2006fg,Campanelli:2006fy}. 
In addition, the
radiation of net linear momentum by a black-hole binary leads to the
recoil of the final remnant hole~\cite{Campanelli:2004zw, Herrmann:2006ks,
Baker:2006vn,Gonzalez:2006md,Herrmann:2007ac,Campanelli:2007ew,
Koppitz:2007ev,Gonzalez:2007hi,Choi:2007eu,Baker:2007gi}. 
This phenomenon can lead to astrophysically important effects 
\cite{Redmount:1989,Merritt:2004xa}.

A non-spinning black-hole binary will emit net
linear momentum parallel to its orbital
plane if the individual holes have  unequal masses. 
However, the maximum recoil in this case (which occurs when the
mass ratio is $q\approx0.36$) is a relatively
small $\sim175\ \KMS$~\cite{Gonzalez:2006md}. 

The first generic simulation of black-hole binaries with unequal
masses and spins was reported in~\cite{Campanelli:2007ew}. These black
holes displayed spin precession and spin flips, and for the first
time, recoil velocities over $400\KMS$, mostly along the orbital
angular momentum direction. It was thus found that the unequal spin
components to the recoil velocity can be much larger than those due to
unequal masses, and that comparable mass, maximally spinning holes
with spins in the orbital plane and counter-aligned would lead to
the maximum possible recoil. This maximum
recoil will be normal to the orbital plane.  Brief studies of this
configuration (with $a/m$ between $0.5$ and $0.8$) were performed
in~\cite{Campanelli:2007ew,Gonzalez:2007hi}.   
In this letter we report on the first systematic study of such
configurations. Consistent and independent recoil velocity
calculations have also been obtained for equal-mass binaries with
spinning black holes that have spins aligned/counter-aligned with the
orbital angular momentum~\cite{Herrmann:2007ac,Koppitz:2007ev}. Recoils
from the merger of non-precessing black-hole binaries have been
modeled in~\cite{Baker:2007gi}.

In~\cite{Campanelli:2007ew} we introduced the following 
heuristic model for the
gravitational recoil of a merging binary.
\begin{eqnarray}\label{eq:empirical}
\vec{V}_{\rm recoil}(q,\vec\alpha_i)&=&v_m\,\hat{e}_1+
v_\perp(\cos(\xi)\,\hat{e}_1+\sin(\xi)\,\hat{e}_2)+\vz\,\hat{e}_z,\nonumber\\
v_m&=&A\frac{q^2(1-q)}{(1+q)^5}\left(1+B\,\frac{q}{(1+q)^2}\right),\nonumber\\
v_\perp&=&H\frac{q^2}{(1+q)^5}\left(\alpha_2^\|-q\alpha_1^\|\right),\nonumber\\
\vz&=&K\cos(\Theta-\Theta_0)\frac{q^2}{(1+q)^5}\left(\alpha_2^\perp-q\alpha_1^\perp\right),
\end{eqnarray}
where $A = 1.2\times 10^{4}\ \KMS$, $B = -0.93$,
$H = (7.3\pm0.3)\times 10^{3}\ \KMS$, $\vec{\alpha}_i=\vec{S}_i/m_i^2$,
$\vec S_i$ and $m_i$ are the spin and mass of
hole $i$, $q$ is the mass ratio of the smaller to larger mass hole,
the index $\perp$ and $\|$
refer to perpendicular and parallel to the orbital angular momentum
respectively, $\hat{e}_1,\hat{e}_2$ are orthogonal unit vectors in the
orbital plane, and $\xi$ measures the angle between the ``unequal mass''
and ``spin'' contributions to the recoil velocity in the orbital plane
(see~\cite{Baker:2007gi} for a similar empirical formula).
The angle $\Theta$ was defined as the angle between the in-plane
component of $\vec \Delta\equiv m({\vec S_2}/m_2 -{\vec S_1}/m_1)$ 
and the infall direction at merger. We determine below that
$K=(6.0\pm0.1)\times 10^4\ \KMS$. We note that the maximum of the recoil
velocity shifts toward equal-mass binaries when spin is present.
For example, in the case where $\vec \alpha_2 = - \vec \alpha_1 =
\vec \alpha$  the maximum recoil occurs for $q=1$ both when 
$\alpha^\perp=0$ for $\alpha\cos(\xi) < 0.0$ and when
$\alpha^\| =0$ for $ \alpha \cos(\Theta - \Theta_0) > 0.07675$.

Current techniques are not accurate enough to measure the spin 
directions of the individual holes at merger. 
Instead, we focus on the angle $\vartheta$ between the initial $\vec \Delta$ 
(which, for our binaries, is parallel to the individual spins
and to the orbital plane)
and the {\it initial} linear (orbital) momenta of the holes.
We test the dependence of the recoil on $\vartheta$ 
by varying the initial spin directions while keeping the initial
puncture positions and momenta fixed. 
In addition, we choose configurations that suppress
$v_m$ and $v_\perp$ and maximize $\vz$.

{\it Techniques:}
We use the puncture approach~\cite{Brandt97b} along with the {\sc
TwoPunctures}~\cite{Ansorg:2004ds} thorn to compute initial data.
In all cases below, we evolve data containing only two
punctures with equal puncture mass parameters, which we denote
by $m_p$.  We evolve these black-hole-binary
data-sets using the {\sc LazEv}~\cite{Zlochower:2005bj} implementation
of the `moving puncture approach' which was independently proposed
in~\cite{Campanelli:2005dd, Baker:2005vv}. 
 In our
version of the moving puncture approach~\cite{Campanelli:2005dd} we
replace the BSSN~\cite{Nakamura87,Shibata95, Baumgarte99} conformal
exponent $\phi$, which has logarithmic singularities at the punctures,
with the initially $C^4$ field $\chi = \exp(-4\phi)$.  This new
variable, along with the other BSSN variables, will remain finite
provided that one uses a suitable choice for the gauge. An alternative
approach uses standard finite differencing of
$\phi$~\cite{Baker:2005vv}.
 We use the
Carpet~\cite{Schnetter-etal-03b} mesh refinement driver to
provide a `moving boxes' style mesh refinement. In this approach
refined grids of fixed size are arranged about the coordinate centers
of both holes.  The Carpet code then moves these fine grids about the
computational domain by following the trajectories of the two black
holes.

We obtain accurate, convergent waveforms and horizon parameters by
evolving this system in conjunction with a modified 1+log lapse and a
modified Gamma-driver shift
condition~\cite{Alcubierre02a,Campanelli:2005dd}, and an initial lapse
$\alpha\sim\psi_{BL}^{-4}$ (here $\psi_{BL} = 1 + m_p/(2 r_1) + m_p /(2 r_2)$,
where $r_i$ is the coordinate distance to puncture $i$).
 The lapse and shift are evolved with
$(\partial_t - \beta^i \partial_i) \alpha = - 2 \alpha K$, $\partial_t
\beta^a = B^a$, and $\partial_t B^a = 3/4 \partial_t \tilde \Gamma^a -
\eta B^a$.
  These gauge conditions require careful treatment of
$\chi$, the inverse of the three-metric conformal factor, near the
puncture in order for the system to remain
stable~\cite{Campanelli:2005dd,Campanelli:2006gf,Bruegmann:2006at}.
As was shown in Ref.~\cite{Gundlach:2006tw}, this choice of gauge
leads to a strongly hyperbolic evolution system provided that the
shift does not become too large.

In Ref.~\cite{Campanelli:2007ew} we presented convergence and consistency tests
for our mesh-refinement code and showed that our code produces
fourth-order accurate waveforms for spinning binaries with spin magnitudes
equal to the spins used in the present configurations and central
resolutions of $M/32$, $M/40$, $M/52$.

{\it Results:}
We evolved the configurations given in Table~\ref{table:ID} with 9
levels of refinement and a finest resolution of $h=M/40$. The outer
boundaries were located at $320M$.
We
measure the gravitational recoil by analyzing the $(\ell,m)$ modes of
$\psi_4$ ($\ell\leq4$) as measured by observers at $r=25M$, $30M$, $35M$, $40M$ and
extrapolating to infinity. We take the error in our measured recoil to be the
differences between a linear and quadratic extrapolation
(in $1/r$) of these measurements. We removed the contribution of the initial
(non-physical) radiation burst (which is typically $\sim20 \KMS$) from
the computed recoil. There are additional (small) errors due to 
our not including the initial non-zero recoil of the system as well as
finite difference errors.
 Note that these configurations all have
$\pi$-rotation symmetry, and consequently, if puncture 1 is located at
$(x_p, y_p, z_p)$ then puncture 2 will be located at $(-x_p, -y_p, z_p)$ 
(note the sign of the $z$ coordinate). Thus these binaries do not
undergo a typical orbital precession, rather the orbital plane itself
moves up and down the $z$-axis.

We obtained the momentum and puncture position initial data parameters
for the SP2 configuration using the 3PN equations for a quasi-circular
binary with period $M\omega = 0.0500$ and the given spin. We
determined the puncture mass parameter by requiring that the ADM mass
be 1. We then rotated the spin, keeping its magnitude constant, to
obtain the parameters for the remaining configurations.

\begin{table}
\caption{Initial data parameters. The punctures are located along the
$x$-axis at $x/M=\pm3.28413$ with momenta $\vec P = \pm(0,0.13355,0)$,
spins $\vec S = \pm (S_x, S_y, 0)$, and puncture mass parameters
$m_p/M=0.430213$. In all cases the specific-spin of the
two holes is $a/m = 0.5$.}
\begin{tabular}{lcccc}
\hline\hline
Config & $\vartheta$ & $S_x$ & $S_y$ & $M_{\rm ADM}/M$ \\
SP2    & 0 & $0$ & $0.12871$ & $1.00001$\\
SPA    & $-\pi/4$ & $0.091013$ & $0.091013$ & $0.99980$\\
SPB    & $\pi/2$ & $-0.12871$ & $0$ & $0.99958$\\
SPC    & $\pi$ & $0$ & $-0.12871$ & $1.00001$\\
SPD    & $\pi/2 + 0.162$ & $-.12703$ & $-0.020760$ & $0.99959$\\
SPE    & $-\pi/2$ & $0.12871$ & $0$ & $0.99958$\\

\hline\hline
\end{tabular} \label{table:ID} 
\end{table} 

\begin{figure}
\begin{center}
\includegraphics[width=2.7in]{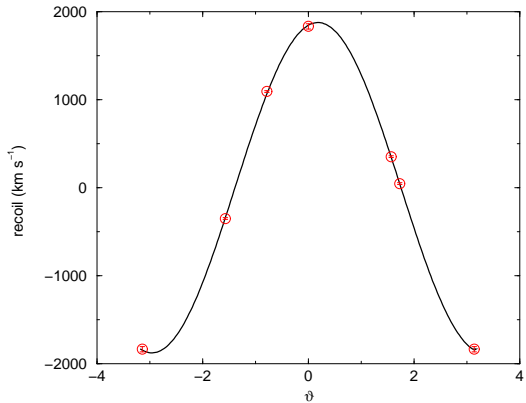}
\caption{The recoil velocity versus angle $\vartheta$ between the initial
individual momenta and spins and a least-squares
fit. Note that the $\vartheta=\pm\pi$ are the same SPC configuration.}
\label{fig:kick_v_angle}
\end{center}
\end{figure}

\begin{figure}
\begin{center}
\includegraphics[width=2.7in]{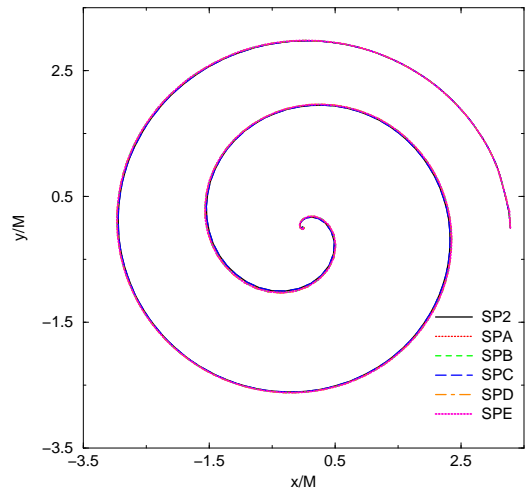}
\caption{The projection of the puncture trajectories (only 1 shown
per configuration) for the 6 configurations. The orbital dynamics
of the binaries are not significantly  affected by the change in spin
directions.}
\label{fig:spn_2dtrack}
\end{center}
\end{figure}

\begin{figure}
\begin{center}
\includegraphics[width=2.7in]{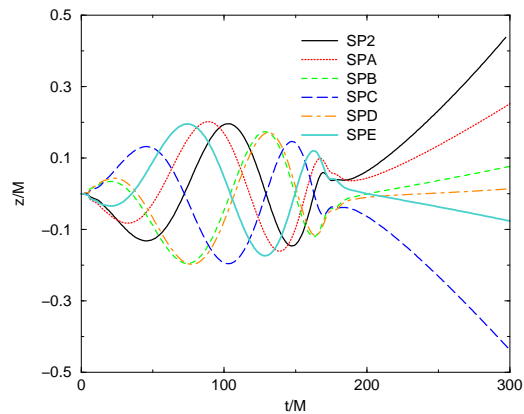}
\caption{The $z$-component of the  punctures  trajectories (only 1 shown
per configuration) versus time for the 6 configurations showing the
dependence  of the orbital plane `precession' and remnant recoil on
the angle of rotation.}
\label{fig:spn_ztrack}
\end{center}
\end{figure}

\begin{figure}
\begin{center}
\includegraphics[width=2.7in]{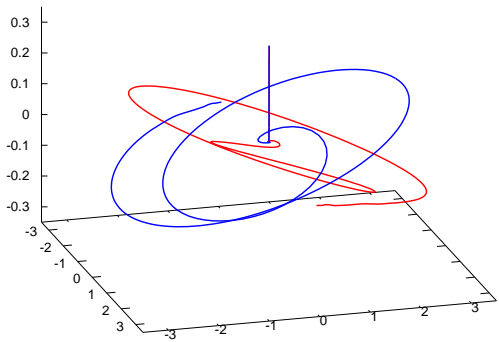}
\caption{
The three-dimensional trajectories of the  punctures
 showing the orbital precession and the final recoil for the SP2
configuration. Note that the scale of the $z$-axis is $1/10$ that
of the $x$ and $y$ axes.}
\label{fig:sp2_3track}
\end{center}
\end{figure}

Table~\ref{table:kick_v_angle} and Fig.~\ref{fig:kick_v_angle}
give the recoil velocity for each configuration. 
A linear least-squares fit for all configurations yields 
$\vz = 1875.87 \cos(\vartheta - 0.183978)$.
Note the very similar values for the radiated energy and angular momenta.
The fact that these values are identical to within $3\%$ in the radiated
energy and to within the errors in the calculation for the radiated
angular momentum, indicates that the binaries have very similar 
orbital dynamics. 
This expectation is further supported by the puncture
trajectories in the $xy$-plane (see Figs.~\ref{fig:spn_2dtrack},
\ref{fig:spn_ztrack}, and~\ref{fig:sp2_3track}) which show essentially identical orbital
trajectories and with an identical number of orbits prior to merger.
Thus we expect that a rotation of the initial spins will lead
to an essentially identical merger but with the spins at merger rotated
by that angle. 
Finally, we note that the PN spin-precession equations~\cite{Kidder:1995zr}
imply that the in-plane spin precession frequency is independent of
spin orientation for our configurations to 1.5 PN order.
Hence our fit of $\vz$ to $v_z \cos(\vartheta - \vartheta_0)$
indicates that $\vz$ varies as $\cos (\Theta - \Theta_0)$ as 
predicted by our empirical formula~(\ref{eq:empirical}).

\begin{table}
\caption{The radiated energy and angular momentum and 
recoil velocity $\vz$ for the configurations in
Table~\ref{table:ID} including predicted velocities based a
least-squares fit of all configurations.}
\begin{tabular}{lllcc}
\hline\hline
Config & $\vz$ ($\KMS$) & $\vz$ (fit) & $E_{\rm rad}/M$ & $J_{\rm rad}/M^2$ \\
SP2    & $1833\pm30$ & $1844$ & $(3.63\pm0.01)\%$& $0.248\pm0.003$\\
SPA    & $1093\pm10$ & $1061$ & $(3.53\pm0.01)\%$ & $0.244\pm0.003$\\
SPB    & $352\pm10$ & $343$ & $(3.57\pm0.01)\%$ & $0.246\pm0.004$\\
SPC    & $-1834\pm30$ & $-1844$ & $(3.63\pm0.01)\%$ & $0.249\pm0.003$\\
SPD    & $47\pm10$ & $41$ & $(3.55\pm0.02)\%$ & $0.245\pm0.005$\\
SPE    & $-351\pm10$ & $-343$ & $(3.57\pm0.02)\%$ & $0.246\pm0.003$\\
\hline\hline
\end{tabular} \label{table:kick_v_angle} 
\end{table} 

{\it Discussion:} In an earlier paper \cite{Campanelli:2007ew} 
we reported the first
results from evolutions of a generic black-hole binary, i.e.\ a binary
containing unequal-mass (2:1) black holes with misaligned spins. 
These results suggested that the recoil velocities of rapidly-rotating
black holes would be dominated by the contribution from the spins.
While the configuration evaluated in that paper was not selected
in order to maximize the recoil, the results were used to estimate the
maximum recoil velocity due to spin, based on an empirical formula,
Eq.~(\ref{eq:empirical}). 
In this {\it Letter}, we have confirmed
our previous estimates with a set of new numerical simulations of
binaries having spins of  equal magnitude but counter-aligned, 
and parallel to the orbital plane. 
We found that these configurations maximize the $\vz$ term in
Eq.~(\ref{eq:empirical}) while setting the remaining terms to zero. 
We confirmed that $\vz$ varies as $K\cos(\vartheta - \vartheta_0)$, 
where $\vartheta$ measures the angle between the {\it initial}
spin and linear momentum vectors.
Based on the fit $\vz = (1876\pm 30) \cos(\vartheta - 0.183978)$ we
determined that $K=(6.0\pm0.1)\times10^{4}\KMS$.
Since the magnitude of the recoil predicted by Eq.~(\ref{eq:empirical})
is proportional to the dimensionless spins ${\vec\alpha_i}$,
our results predict maximum recoil velocities of $\sim 4000\ \KMS$
in the case of maximally-spinning holes with counter-aligned spins.

A post-merger recoil velocity of $\sim 4000\ \KMS$ is large enough to eject
a black hole from the center of even the most massive elliptical
galaxies~\cite{Merritt:2004xa}.
Hence, our results strengthen the conclusion, already reached in several
recent papers \cite{Campanelli:2007ew,Gonzalez:2007hi,Baker:2007gi}
that radiation recoil is capable of completely removing supermassive black
holes (SMBHs) from their host galaxies.
Computing the probability of such an extraordinary event will require a more
extensive set of numerical simulations that characterize the dependence of
$V_{\rm recoil}$ on spin direction for generic binaries, with arbitrary spin
orientations and mass ratios.
Here, we note the strong predicted dependence of $V_{\rm recoil}\sim q^2$ on
mass ratio which implies a ``maximum'' recoil velocity of $\sim
0.3^2\times 4000\ \KMS\approx 400\ \KMS$ even for a ``major merger'' with
$m_2/m_1\approx 1/3$.
In addition, the root-mean-square recoil velocity for randomly oriented 
spins in the plane
is reduced by an additional factor of $\sqrt{2}$.
Our results are therefore not inconsistent with the observed fact that SMBHs
are apparently ubiquitous components of luminous galaxies.
A galaxy in which the SMBH has been permanently removed will appear similar
to a more ordinary galaxy except for a larger stellar core, or ``mass
deficit,''
due to heat input from the ejected hole.
In fact, observed mass deficits are sometimes $\sim $ a few times larger
than predicted on the basis of binary SMBH models \cite{merritt-06}.
The detection (or non-detection) of a SMBH in such a galaxy via stellar
absorption line features will be difficult however due to its low central
surface brightness.

We thank Erik Schnetter for valuable discussions and providing {\sc
Carpet}.  We thank Marcus Ansorg for providing the {\sc TwoPunctures}
initial data thorn and Johnathan Thornburg for providing {\sc
AHFinderDirect}.  We gratefully acknowledge NSF for financial support
from grant PHY-0722315.  D. M. was supported by grants AST-0420920 and
AST-0437519 from the NSF and grant NNG04GJ48G from NASA.
Computational resources were provided by Lonestar cluster at TACC.
   
\bibliographystyle{apsrev}
\bibliography{references}

\end{document}